\documentclass[11pt,preprint]{aastex}
\usepackage{epsfig,color}
\usepackage{mathrsfs}
\usepackage{graphicx}
\usepackage{bm}
\usepackage{amsmath}

\shorttitle{Short Term Variability of S5 0716+714} \shortauthors{Feng et al.}

\begin{document}

\title{Quasi-simultaneous Spectroscopic and Multi-band Photometric Observations of Blazar S5 0716+714 during 2018-2019}

\author{Hai-Cheng Feng\altaffilmark{1,2,3,\bigstar},
H. T. Liu\altaffilmark{1,3,4, \bigstar},
J. M. Bai\altaffilmark{1,3,4, \bigstar},
L. F. Xing\altaffilmark{1,3,4},
Y. B. Li\altaffilmark{1,2,3},
M. Xiao\altaffilmark{5},
Y. X. Xin\altaffilmark{1,3,4}}

\altaffiltext{1} {Yunnan Observatories, Chinese Academy of Sciences,
Kunming 650011, Yunnan, P. R. China.}

\altaffiltext{2} {University of Chinese Academy of Sciences, Beijing
100049, P. R. China.}

\altaffiltext{3} {Key Laboratory for the Structure and Evolution of
Celestial Objects, Chinese Academy of Sciences, Kunming 650011,
Yunnan, P. R. China}

\altaffiltext{4} {Center for Astronomical Mega-Science, Chinese Academy of Sciences, 20A Datun Road, Chaoyang District, Beijing, 100012, P. R. China}

\altaffiltext{5} {Key Laboratory for Particle Astrophysics, Institute of High Energy Physics, Chinese Academy of Sciences, 19B Yuquan Road, Beijing, 100049, P. R. China}

\altaffiltext{$^{\bigstar}$}{Corresponding authors: Hai-Cheng. Feng, e-mail: hcfeng@ynao.ac.cn,
H. T. Liu, e-mail: htliu@ynao.ac.cn, J. M. Bai, e-mail: baijinming@ynao.ac.cn}

\begin{abstract}
  In order to study short timescale optical variability of $\gamma$-ray blazar S5 0716+714, quasi-simultaneous spectroscopic and multi-band photometric observations were performed from 2018 November to 2019 March with the 2.4 m optical telescope located at Lijiang Observatory of Yunnan Observatories. The observed spectra are well fitted with
  a power-law $F_{\lambda}=A\lambda ^{-\alpha}$ (spectral index $\alpha >0$). Correlations found between $\dot{\alpha}$, $\dot{A}$, $\dot{A}/A$, $\dot{F_{\rm{\lambda}}}$, and $\dot{F_{\rm{\lambda}}}/F_{\rm{\lambda}}$ are consistent with the trend of bluer-when-brighter (BWB). \textbf{The same case is for colors, magnitudes, color variation rates, and magnitude variation rates of photometric observations.} The variations of $\alpha$ lead those of $F_{\rm{\lambda}}$. Also, the color variations lead the magnitude variations. The observational data are mostly distributed in the I(+,+) and III(-,-) quadrants of coordinate system. Both of spectroscopic and photometric observations show BWB behaviors in S5 0716+714. The observed BWB may be explained by the shock-jet model, and its appearance may depend on the relative position of the observational frequency ranges with respect to the synchrotron peak frequencies, e.g., at the left of the peak frequencies. \textbf{Fractional variability amplitudes are $F_{\rm{var}}\sim 40\%$ for both of spectroscopic and photometric observations. Variations of $\alpha$ indicate variations of relativistic electron distribution producing the optical spectra. }

\end{abstract}

\keywords{BL Lacertae objects: general --- BL Lacertae objects: individual (S5 0716+714) --- galaxies: active ---
galaxies: photometry --- techniques: spectroscopic}

\section{Introduction}
\textbf{Blazars are a subclass of active galactic nuclei (AGNs) }and usually exhibit extreme variability in the whole electromagnetic spectrum \citep[e.g.,][]{Ul97}. \textbf{Depending on the rest-frame equivalent widths (EWs), blazars
can be divided into BL Lacertae objects (BL Lacs) and flat-spectrum radio quasars (FSRQs). The EWs of BL Lacs and FSRQs
are $<$ 5 $\mathring{\rm{A}}$ and $>$ 5 $\mathring{\rm{A}}$, respectively \citep[e.g.,][]{Gh11,GT15}.} Generally, the continuum radiation of BL Lacs is believed to be relativistically boosted along the line of sight by relativistic jets
with small viewing angles \citep[e.g.,][]{UP95,Ul97} and shows observational characteristics, such as featureless optical spectra, strong non-thermal emission, and high polarization, etc. There are two peaks in broadband spectral energy distributions (SEDs) of \textbf{blazars} \citep[e.g.,][]{Gh98,Ul97}. Their low and high energy peaks are located around \textbf{from infrared-}optical-ultraviolet (UV) \textbf{to} X-ray bands and \textbf{around MeV-GeV-TeV} $\gamma$-ray bands, respectively. The low energy peak is the synchrotron radiation from relativistic electrons in the relativistic jets and
the high energy peak, the $\gamma$-ray emission, is generally interpreted as the inverse-Compton (IC) scattering of the synchrotron soft photons \textbf{for blazars} and the external soft photons \textbf{for FSRQs} by the same electron distribution that radiates the synchrotron photons \citep[e.g.,][]{Ul97,Gh98,CG08,Ta10,Ne12,Zh12,MS16,Zh17}.

Various variability timescales, e.g., from minutes to decades, have been found in most BL Lacs and these timescales
can help us to investigate the properties of radiation region \citep[e.g.,][]{Xi99,Xi02,Xi05,Co15,Li15,Wi15,Fe17,Li19}.
The variability timescales are usually divided into three classes: the timescales less than one night are regarded as intra-day variability (IDV) or micro-variability \citep[e.g.,][]{WW95,Fa14}; the timescales from days to a few months are short-term variability (STV) \citep[e.g.,][]{Li17}; and the timescales larger than several months are known as long-term variability (LTV) \citep[e.g.,][]{Da15}. Different variability timescales may be originated from different emission
regions. Thus, we can study different radiation mechanisms via variability with different timescales. Furthermore, the
flux variability often follows different spectral behavior and the correlation between the variability of flux and
spectral index (or magnitude and color) will shed light on the physical processes of radiation for BL Lacs. A common phenomenon has been found in most BL Lacs. The bluer spectral index usually arises at the brighter phase in most BL Lacs \citep[e.g.,][]{Vi04,Bo12}, i.e., bluer-when-brighter (BWB). The BWB trend is often regarded as evidence of shock-in-jet model \citep[e.g.,][]{MG85,Gu08,Bo12}. However, many observations do not show any correlation between colors and magnitudes \citep[e.g.,][]{Ag16,Ho17}, or show only weak correlations \citep[e.g.,][]{Wi15}. The discrepancy of the color-magnitude correlations is a crucial issue that can help us to understand more detailed radiation properties in jets.

S5 0716+714 is a typical BL Lac object at a redshift of $0.31\pm0.08$ \citep{Ni08}. It was first discovered by \citet{Ku81}
and was widely studied on the whole electromagnetic spectrum \citep[e.g.,][]{Os06,Ab10,Hu14,Da15,Ga15,Fe17,Ho17,Sa17,Li19}.
It is one of the most active and bright BL Lacs in the optical band and shows a completely featureless spectrum \citep[e.g.,][]{Bi81,Da13}. A number of groups have focused on the broadband photometric study of S5 0716+714 in
the optical regime. Almost, all of them have found the variability with timescales from mins to years \citep[e.g.,][]{Ne02,Hu14,Da15,Ag16,Ho17,Li17}. Many studies reported high IDV duty cycles \citep[DCs;][]{WW95}, i.e.,
DCs $\geq 70\%$ for S5 0716+714. Variation amplitudes are larger than 0.05 mag for 80\% of 52 nights \citep{Ne02}. \citet{Hu14} gave a DC of 83.9\% on 42 nights. \citet{Ag16} obtained a DC of $\sim$90\% by 23 night observations. The probability of variability in S5 0716+714 is nearly daily. The various (strong, weak, or non) BWB trends have been also reported in many observations. \citet{Dai13} found that the source exhibited strong BWB chromatism in LTV, STV, and IDV. \citet{Hu14} showed strong and mild BWB trends on IDV and STV, respectively. \citet{Ag16} did not found any correlations between colors and magnitudes. Recently, \citet{Ho17} reported an outburst state during 2012 and they found both BWB chromatism and weak BWB trend in most nights. However, in few nights, the data did not show any correlations between
colors and magnitudes. The observational characteristics mentioned above indicate that S5 0716+714 is a natural laboratory for studying the radiation properties of BL Lacs.

\textbf{Almost all of the previous studies only used a few of broadband photometric observations. Bandwidths of broadband filters are usually larger than 1000 $\mathring{\rm{A}}$ and different filters have different bandwidths. Therefore, the relationship between brightness and spectral behavior is only roughly studied. The broad bandwidths might also influence
the relationship during some phases (e.g., might decrease the correlation coefficient during weak phases). Moreover, the adjacent bands will partly overlap each other, which will further influence the correlation between the brightness and spectral behavior.} In order to investigate the relationships of index--flux, index variability--flux variability, color--magnitude, and color variability--magnitude variability, and shed some light on the radiation processes of BL Lacs, we simultaneously monitored S5 0716+714 with spectroscopic observations and broadband photometry. \textbf{The spectral data can provide the light curves (LCs) at narrow enough wavelength coverage which allow us to study the above relationships in details. Besides, comparing photometric LCs to spectral integral LCs will help us probe the effect of bandwidth.} The correlations of variability among different bands and different wavelength ranges could also help us to limit the relative location of radiation.

In Section 2, we describe the detailed information of observations and data reductions. The results and our analyses are presented in Section 3. Finally, discussion and conclusion are presented in Section 4.

\section{Observations and Data Reduction}

All the spectroscopic and photometric observations of S5 0716+714 were carried out with the 2.4 m alt-azimuth
telescope, which is located at Lijiang Observatory of Yunnan Observatoris, Chinese Academy of Science. The longitude, latitude, and altitude of the observatory are 100$^{\circ}$01$^{\prime}$48$^{\prime\prime}$, 26$^{\circ}$42$^{\prime}$42 $^{\prime\prime}$, and 3193 m, respectively. From mid-September to May, the observatory is dry and most nights are
clear. The average seeing of the telescope obtained by the full width at half maximum (FWHM) of stars is $\sim$ 1$^{\prime\prime}_{\cdot}$5 \citep[e.g.,][]{Du14}. For the 2.4 m telescope, the pointing accuracy is about
2$^{\prime\prime}$, and the closed-loop tracking accuracy is better than 0$^{\prime\prime}_{\cdot}$5 hr$^{-1}$. In 2010,
the telescope was mounted with an Yunnan Faint Object Spectrograph and Camera (YFOSC) at Cassegrain focus. This is an all-purpose CCD for low/medium dispersion spectroscopy and photometry. The CCD can keep low readout noise under high
readout speed, which benefits from all-digital hyper-sampling technology. During our observations, the readout noise
and gain are 9.4 electrons and 0.35 electrons/ADU, respectively. The CCD chip covers a field of view (FOV) of 9$^{\prime}_{\cdot}$6$\times$ 9$^{\prime}_{\cdot}$6 with 2048 $\times$ 4096 pixels, and the pixel scale is 0$^{\prime\prime}_{\cdot}$283 pixel$^{-1}$. YFOSC can quickly switch from photometry to spectroscopy ($\leq$ 1 s),
and we can also choose the binning mode to reduce the photometric readout time. The detailed parameters of
the telescope and YFOSC were described in \citet{Wa19}.

The monitoring campaign started in 2018 November and spanned $\sim$ 106 days. For most clear dark or grey nights, we basically performed photometric and spectroscopic observations of S5 0716+714 within 10 minutes. Thus, the photometry
and spectroscopy can be considered to be quasi-simultaneous. During our observations, we successfully obtained the photometric data in 42 nights and spectral data in 47 nights. The cadence of spectroscopy is $\sim$ 2.08 days.
The complete observation information is listed in Table 1.

\subsection{Photometry}

The photometric observations were performed using Johnson $BV$ and Cousins $RI$ filters. In order to obtain the accurate magnitude calibration of the target, we always set several comparison stars in the observed FOV. The comparison stars
were presented in \citet{Vi98}, who have calibrated the magnitudes in the $BVR$ bands. We found that star2, star3, star5, and star6 are closest to the target \citep[see Figure 3 in][]{Vi98}. Besides, the four comparison stars were also used in \citet{Gh97}, who gave the data of the $I$ band. Thus, these stars are selected as comparison stars in our observations.
The magnitude of S5 0716+714 is calibrated as follows:
\begin{equation}
Mag = -2.5\; \log\frac{1}{N}\sum_{i}^{N}10^{-0.4(M^{i}_{\rm{std}} + M_{\rm{o}} - M_{i})},
\label{equation 1}
\end{equation}
where $N$ is the number of comparison stars, $M^{i}_{\rm{std}}$ is the standard magnitude of the $i$th comparison star,
and $M_{\rm{o}}$ and $M_{i}$ are the instrumental magnitudes of the target and the $i$th comparison star, respectively.
\textbf{Figure 1} shows the calibrated LCs of S5 0716+714. The calibration errors include two components. The first is
the Poisson errors of the target and comparison stars, and it can propagate through Equation (1). The second is from the systematic uncertainties which might be caused by the phase of the moon, weather condition, etc. We calibrated one of the comparison stars (star3) using Equation (1) and the variability of the star can be regarded as the systematic error. The different band calibrated magnitudes of S5 0716+714 and star3 are listed in Tables 3--6. The systematic error is calculated by
\begin{equation}
\sigma_{\rm{sys}} = Mag_{3} - \overline{Mag_{3}},
\label{equation 2}
\end{equation}
where $Mag_{3}$ is the calibrated magnitude of star3. Finally, the errors are $\leq 1\%$ in most nights. The errors
are also listed in Tables 3--6.

All the photometric data were reduced using standard Image Reduction and Analysis Facility (IRAF) software. After
the bias and flat-field corrections, we extracted the instrumental magnitudes of the target and comparison stars with
different apertures. To avoid the contamination of the host galaxy mentioned in \citet{Fe17}, we tested two different apertures: dynamic apertures (several times FWHM) and fixed apertures. For each type of aperture, we chose 10 different apertures. The aperture radii of fixed apertures and dynamic apertures are 1$^{\prime\prime}_{\cdot}$5--8$^{\prime\prime}_{\cdot}$0 and 1.3--3.5 $\times$ FWHM, respectively. The results are
almost the same in different apertures. However, the best signal-to-noise ratio (S/N) could be obtained with the
aperture radius of 6$^{\prime\prime}_{\cdot}$0, and we adopted the photometry under this aperture as the final result.

\subsection{Spectroscopy}

Considering the featureless spectra of BL Lacs, the spectroscopic observations were carried out with Grism 3, which
provides a relatively low dispersion (2.93 $\mathring{\rm{A}}$ pixel$^{-1}$) and wide wavelength coverage (3400--9100 $\mathring{\rm{A}}$). We found that the spectrum of Grism 3 might be slightly contaminated by the 2nd order spectrum
as wavelength is longer than $\sim$ 7000 $\mathring{\rm{A}}$, and the 2nd order spectrum is $\sim 5\%$ times intensity
of the 1st order spectrum. To avoid the effect of the 2nd order spectrum, we use a UV-blocking filter which cuts off
at $\sim$ 4150 $\mathring{\rm{A}}$. Thus, the secondary spectrum will be rejected shorter than $\sim$ 8300 $\mathring{\rm{A}}$. The final spectra cover the observed-frame of 4250--8050 $\mathring{\rm{A}}$. To improve the flux calibration, we simultaneously put the target and star3 in the long slit. This method was used widely \citep[e.g.,]
[]{Ka20, Du14,Lu16}, and can obtain the relatively high quality spectra even in poor weather. To minimize the effects
of seeing, we use a wide slit with a projected width of 5$^{\prime\prime}_{\cdot}$05. For each night, we also observe 
a spectrophotometric standard star, which can calibrate the absolute fluxes of the target and comparison star.

The raw spectral data are also reduced with IRAF. After correcting the bias and flat-field, we calibrate the wavelength
of two-dimensional spectral image using standard helium and neon lamps. We extract the spectra of the target and star3
after removing the cosmic-rays. The extraction aperture radius is 21 pixels ($\sim$ 5$^{\prime\prime}_{\cdot}$943),
nearly the same with photometry. We calibrate the absolute fluxes of the target and star3 using the spectrophotometric standard star. Note that miscentering of the object in slit will cause the shift of wavelength and then will influence
the calibration of flux. We correct the shift by the absorption lines from 6400--7100 $\mathring{\rm{A}}$. In the end, we re-calibrate the spectra using the template spectrum of comparison star. The template spectrum is obtained by averaging 
the spectra of star3, which are observed in the nights with good weather conditions. The absorption lines of atmosphere are also corrected by the comparison star. Figure 2 is the mean spectrum and an individual spectrum.

We bin individual spectra to obtain the spectroscopic LCs, and the bin width is 50 $\mathring{\rm{A}}$. The flux and
error of each bin are obtained by the mean and standard deviation of the fluxes in the corresponding bin, respectively.
We find that the LC of each bin is nearly the same with each other, and then only 6 bins with the centers of 4425, 5125, 5825, 6525, 7225, and 7925 $\mathring{\rm{A}}$ are used for analysis. The 6 bins are denoted in the top panel of Figure 2,
and the relevant LCs are shown in Figure 3.

\subsection{Fractional Variability Amplitude and Spectral Index}
The variability amplitude of each light curve is calculated by the root-mean-square (RMS) fractional variability amplitude $F_{\rm{var}}$ \citep[e.g.,][]{Ro97,Ed02,Va03}. The fractional variability amplitude $F_{\rm{var}}$ is defined as
\begin{equation}
   F_{\rm{var}}=\frac{\sqrt{S^2-<\sigma^2_{\rm{err}}>}}{<F>},
   \label{equation 3}
\end{equation}
where $S^2$ denotes the total variance for the $N$ data points in a light curve, $<F>$ is the mean flux of the light curve, and $<\sigma^2_{\rm{err}}>$ denotes the measured mean square error of the $N$ data points:
\begin{mathletters}
  \begin{eqnarray}
     S^2 = \frac{1}{N-1}\sum^{N}_{\rm{i=1}}(F_{\rm{i}}-<F>)^2,\\
     <F> = \frac{1}{N}\sum_{\rm{i}}^{N}F_{\rm{i}},\\
     <\sigma^2_{\rm{err}}> = \frac{1}{N}\sum^{N}_{\rm{i=1}}\sigma^2_{\rm{err,i}}.
  \end{eqnarray}
  \label{equation 4}
\end{mathletters}
\citet{Ed02} gave the error $\sigma_{F_{\rm{var}}}$ on $F_{\rm{var}}$:
\begin{equation}
  \sigma_{F_{\rm{var}}}=\frac{1}{F_{\rm{var}}}\sqrt{\frac{1}{2N}}\frac{S^2}{<F>^2}.
  \label{equation 5}
\end{equation}

First, we convert all the photometric data to flux. Then, we measure both spectral and photometric variability amplitude. The variability amplitudes of different LCs are listed in Table 2. The spectral indices and amplitudes of S5 0716+714 are obtained by fitting the spectra via a power-law ($f_{\lambda} = A \lambda^{-\alpha}$). Figure 2 shows the best fit to the mean spectrum and individual spectrum. The variability of spectral index is shown in the left top panel of Figure 3.

\section{Results and Analysis}

During our observations, the amplitudes of variability are $\sim$40\%, calculated from Equation (3). The photometric
and spectroscopic results of $F_{\rm{var}}$ are consistent with each other and show that the variability amplitudes of
S5 0716+714 in the blue side are consistent with those in the red side as considering the relevant uncertainties (see
Table 2). The band widths of the filters are hundreds to thousands angstroms, and the variability amplitudes of photometry are the average results of broad bands. The width of the spectral bin is much narrow than the filter band width. Though
there are the differences between the photometric and spectroscopic bandwidths and bins, the very close wavelength coverage should result in their consistent $F_{\rm{var}}$ for the photometric and spectroscopic observations.

To compare the variability of different bands, we shift each photometric LC to the same level depending on the magnitude
at JD$\sim$ 2458545.12 (the median magnitude of each LC). Figure 4 shows the shifted results. In addition to the differences of the variability amplitudes of valleys, the LCs of different filters are nearly the same as each other. We measure the time delay among different photometric LCs. However, we do not find any reliable time lags. The result of interpolated cross-correlation function \citep[ICCF,][]{WP94, Wa16} between $I$ and $B$ is shown in the right bottom panel of Figure 5. We also test the time delays between the photometric and spectroscopic LCs (see Figure 3), and the LCs are consistent with each other. Therefore, the variability in different wavelength ranges should originate from the same region, and the variations of brightness might cause the changes of color and spectral index.

We find that the variability of different colors is similar to that of the photometric LCs (see Figure 5). The spectral index variability is also similar to that of the LC of each bin (see Figure 3). We test correlations between different colors and different magnitudes. Figures 5 and 6 show the test results. The results indicate that the bluer spectra usually occur at brighter phases, i.e., BWB. The Spearman rank correlation between $B-I$ and $B$ is significant, and other colors are also correlated with $B$. The BWB trend was often found in S5 0716+714 (see Section 1) and can be explained with a shock-in-jet model. The larger variability amplitude is inclined to occur at the shorter wavelength. Thus, the BWB tend
will be more significant when the interval of effective wavelengths between two bands is larger. As mentioned in Section 1, there are some groups which do not find any correlations between the colors and magnitudes. The discrepancy might be caused by the following reasons:

1. For some extended sources, the contamination of the host galaxies might lead to some fake variability because of the change of seeing \citep[e.g.,][]{Fe17,Fe18}. As a result, the observed correlation of the color-magnitude may not be related to the radiation processes. For point sources, the strong host galaxies may dilute the variability amplitudes of AGNs, and then influence the correlation between flux and spectral index, especially during the weak states. S5 0716+714 is a point source and its host galaxy is more than four times darker than the target itself \citep{Ni08}. Thus, the discrepancy should not be caused by the effect of the host galaxy.

2. The accuracy of photometry may also influence the variability of colors. Most photometric studies are based on the
small telescopes ($\leq$ 1 m). For most BL Lacs, the typical variability amplitudes of colors are $\sim$0.05 mag \citep[e.g.,][]{St06, Hu14, AG15}. Furthermore, the variability amplitudes might be less than 0.02 mag for some adjacent
bands. When the photometric accuracy is larger than 0.01 mag, the color-magnitude correlations will be seriously affected. The accuracy of our photometric measurement is less than 1\% in most nights. So, the adjacent bands can show the mild BWB trends (see Figure 6).

During our observations, the entire data show that the BWB trend exists in S5 0716+714. The S/N and sampling frequency of the data are high enough. Therefore, the BWB trend may be an intrinsic phenomenon of the source. The color-magnitude data roughly obey the BWB trend, but the data scatter is visible as well (see Figure 6). The variability of flux density and spectral index are similar to each other (see Figures 3 and 5). Thus, the variability rate of flux might influence the variability of spectral index. Another possibility is that the variability of flux density and spectral index may result from changes of relativistic electron distribution emitting the observed photons and may have a correlation between the relevant variability rates. Thus, we test whether a correlation exists between the variability rates of flux density and spectral index. The sampling of observational data is nearly homogeneous and the variability rates of flux density $F_{\lambda}$, spectral index $\alpha$, and spectral amplitude $A$ are defined as
\begin{mathletters}
  \begin{eqnarray}
  \dot{F_{\rm{\lambda}}} = \frac{F^{\rm{\lambda}}_{\rm{i+1}} - F^{\rm{\lambda}}_{\rm{i}}}{T_{\rm{i+1}} - T_{\rm{i}}},\\
  \dot{\alpha} = \frac{\alpha_{\rm{i+1}} - \alpha_{\rm{i}}}{T_{\rm{i+1}} - T_{\rm{i}}},\\
  \dot{A}= \frac{A_{\rm{i+1}}-A_{\rm{i}}}{T_{\rm{i+1}} - T_{\rm{i}}},
  \end{eqnarray}
\end{mathletters}
where$F^{\rm{\lambda}}_{\rm{i}}$, $\alpha_{\rm{i}}$, and $A_{\rm{i}}$ are the flux density, spectral index, and spectral amplitude observed at the time series $T_{\rm{i}}$, respectively. Figure 7 shows a positive correlation between $\dot{\alpha}$ and $\dot{F_{\rm{\lambda}}}$ for Bin1. The most of data of BWB behavior is distributed in I and III quadrants of coordinate system (see Figure 7). At the same time, there are strong positive correlations of $\dot{\alpha}$--$\dot{A}$ and $\dot{F_{\rm{\lambda}}}$--$\dot{A}$ for Bin1 (see Table 8). Also, there is a correlation between the variability rates of $B$ and $B - I$ and nearly the data of BWB behavior are distributed in I and III quadrants (see Figure 8). Hereafter, spectral index-flux density and color-magnitude relations are called as "color-brightness" relation. These correlations indicate that the variability rates of color and brightness are likely dominated by the cooling and accelerating processes of the relativistic electrons that generate the observed photons and the relevant variability. In Equations (6a)--(6c), the variability rates are calculated from the differences of adjacent data points. The adjacent data points may be considered 
to originate from the same flare.

In order to compare the color-magnitude variability rate correlations with the spectral index-flux density variability
rate correlation, a relative variability rate of flux density is defined as
\begin{equation}
  \frac{\dot{F_{\rm{\lambda}}}}{F_{\rm{\lambda}}} =\left(\frac{F^{\rm{\lambda}}_{\rm{i+1}}}{F^{\rm{\lambda}}_{\rm{i}}}-1\right) \frac{1}{T_{\rm{i+1}} - T_{\rm{i}}}.
\end{equation}
If the flux variability is mainly caused by the variability of spectrum $F_{\rm{\lambda}}=A\lambda ^{-\alpha}$,  $\dot{F_{\rm{\lambda}}}/F_{\rm{\lambda}}$ will be a function of $\dot{A}/A$ and $\dot{\alpha}$, where
\begin{equation}
  \frac{\dot{A}}{A} =\left(\frac{A_{\rm{i+1}}}{A_{\rm{i}}}-1\right) \frac{1}{T_{\rm{i+1}} - T_{\rm{i}}}.
\end{equation}
 The observational data of $\dot{F_{\rm{\lambda}}}/F_{\rm{\lambda}}$ and $\dot{\alpha}$ can be linearly fitted with $\dot{F_{\rm{\lambda}}}/F_{\rm{\lambda}}=B+C\dot{\alpha}$. The Spearman's rank correlation test shows a strong
 positive correlation between $\dot{F_{\rm{\lambda}}}/F_{\rm{\lambda}}$ and $\dot{\alpha}$ (see Table 8), and the BWB
 data of S5 0716+714 are mostly distributed in I and III quadrants (see Figure 9). $B$ is almost close to zero, and
 $C=3.29\pm0.23$. The Spearman's rank correlation analyses show strong positive correlations of $\dot{\alpha}$--$\dot{A}/A$
 and $\dot{A}/A$--$\dot{F_{\rm{\lambda}}}/F_{\rm{\lambda}}$ (see Table 8). Since three correlations exist among $\dot{\alpha}$, $\dot{F_{\rm{\lambda}}}/F_{\rm{\lambda}}$, and $\dot{A}/A$, there should be a correlation like as $\dot{F_{\rm{\lambda}}}/F_{\rm{\lambda}}(\dot{A}/A,\dot{\alpha})$ (see Figure 10). In fact, there is a correlation
 among $\dot{\alpha}$, $\dot{A}/A$, and $\dot{F_{\rm{\lambda}}}/F_{\rm{\lambda}}$ at the confidence level of $>99.99\%$, $\dot{F_{\rm{\lambda}}}/F_{\rm{\lambda}}=0.001 + 0.012\dot{A}/A+ 1.839\dot{\alpha}$. Since I and III quadrants in Figures 7--9 correspond to the BWB, II and IV quadrants in Figures 7--9 should correspond to redder-when-brighter (RWB), which likely have $F_{\rm{\lambda}}=D\lambda^{\alpha}$ in the optical band. Spectroscopic and photometric observations
 show consistent BWB trends in the color-brightness diagrams (Figures 7--9).

\textbf{In order to confirm the Spearman's rank test results listed in Table 8, a Monte Carlo (MC) simulation is used
to reproduce these parameters presented in Table 8. For each pair of these parameters, each data array generated by
the MC simulation is fitted with the SPEAR \citep{Pr92} and the fitting gives the relevant $r_{\rm{s}}$ and $P_{\rm{s}}$, the Spearman's rank correlation coefficient and the p-value of hypothesis test. Considering the errors of X and Y and assuming Gaussian distributions of X and Y, $r_{\rm{s}}$ and $P_{\rm{s}}$ distributions are generated by the SPEAR fitting
to the data of X and Y from $10^4$ realizations of the MC simulation. Averages, $r_{\rm{s}}$(MC) and $P_{\rm{s}}$(MC),
are calculated by the $r_{\rm{s}}$ and $P_{\rm{s}}$ distributions, respectively. Standard deviations of these two distributions are taken as the relevant uncertainties of $r_{\rm{s}}$(MC) and $P_{\rm{s}}$(MC) (see Table 8). These
results given by the MC simulation confirm the ordinary Spearman's rank test results listed in Table 8. Thus, these correlations will be reliable.}

\section{Discussion and Conclusion}
We also test the BWB trend using the bin flux and spectral index (see Figure 11). This BWB trend is slightly different 
from that of color-magnitude. The data are fitted with a fifth-order polynomial and a monotonically increasing trend 
appears in Figure 11. Figure 11 shows that the BWB trend might depend on the brightness. Thus, the relevant radiation 
of the BWB at least includes two components: one component is caused by the propagation of shocks in jet; another 
component is the underlying radiation which is not related with the shock process. If the particles in jet is homogeneous, the variability of BL Lacs should be caused by the disturbance of magnetic field \citep[e.g.,][]{Ch15}, the precession of jet \citep[e.g.,][]{CK92}, the inhomogeneous region of jet, etc. The variations of the underlying radiation of jet may not cause the change of spectral index. But, during a weaker phase the BWB trend caused by the shock will be more significant and during a brighter phase the underlying radiation might dilute the BWB trend. This possibility needs more observation evidences to test. \textbf{There is a possible discrepant point, the one at the lower left quarter in Figure 11, that might affect the fitting result. We exclude this point and re-fit the rest data. The result is very similar to the previous one. The reason that one flux may correspond to several $\alpha$ values is that the spectrum fitting includes two parameters $A$ and $\alpha$. Different $A$ and $\alpha$ combinations may give the same flux. This will result in the data point scatter of BWB for both of spectroscopic and photometric observations. Though the dispersion of $\alpha$ exists, the BWB trend roughly holds (see the best fittings in Figure 11).}

The BWB behavior is observed in our monitoring epoch with the 2.4 m optical telescope located at Lijiang Observatory
of Yunnan Observatories. The BWB behavior can be explained by the shock-jet model. A relativistic shock propagating
down a jet will accelerate electrons to higher energies, where the shock interacts with a nonuniform region of high
magnetic field and/or electron density, likely observed to be knots in jets. The shock acceleration will cause radiations
at different frequencies being produced at different distances. The synchrotron peak frequency depends on the relativistic electron distribution and the magnetic field, i.e., the distances behind the shock front, and the radiation cooling will make the synchrotron radiation peak decrease at the intensity and the frequency. Thus, frequency dependence of the duration of a flare corresponds to an energy-dependent cooling length behind the shock front, which will cause colour variations
in blazars. \citet{Pa07} proposed that the observations during early rising phase of the flux will give a bluer colour
while those taken during later phases of the same flare will show more enhanced redder fluxes. The synchrotron peak of
SED of S5 0716+714 is located very close to the optical wavelengths, and the corresponding broadband SED can be well explained by the synchrotron self-Compton (SSC) and the external radiation Compton (ERC) models, where the SSC soft photons are the synchrotron photons and the ERC soft photons in the IC scattering are emission from a broad-line region (BLR) and/or infrared (IR) emission from a dust torus \citep[e.g.,][]{Li14}. No emission lines were detected in the IR, optical, and UV spectra of S5 0716+714 \citep{Ch11,Sh09,Da13}, and this may from the fact that thermal emission from accretion disk is not found in multiwavelength SED of S5 0716+714 \citep[e.g.,][]{Li14}. The ionizing radiation from accretion disk is so weak that broad emission lines are not observable, even though a BLR exists in S5 0716+714. Also, the dust emission is not observable because of very weak emission of accretion disk, even though a dust torus exists in S5 0716+714.

The observational frequency band is at the left of the synchrotron radiation peak because $F_{\rm{\lambda}}=A\lambda ^{-\alpha}$ ($\alpha >0$). This corresponds to the BWB behavior data in the I(+,+) and III(-,-) quadrants of coordinate system. If the observational frequency band is at the right of the synchrotron radiation peak, we may have $F_{\rm{\lambda}} =D\lambda ^{\alpha}$ ($\alpha >0$). This may correspond to the RWB behavior in the II(-,+) and IV(+,-) quadrants of coordinate system. The first case is observed in our observations and the second one is not observed in our observations. The BWB trends usually arise in most BL Lacs \citep[e.g.,][]{Vi04,Bo12}, and this is probably because that the synchrotron peaks are at optical-UV-X-ray bands for most BL Lacs and that the optical observations are usually at the left of the synchrotron radiation peak. No or only weak BWB trends are observed in many observations \citep[e.g.,][]{Wi15,Ag16,Ho17}, and this may result from that the observational frequency ranges span the synchrotron peak frequencies. Also, the optical variability may be produced by a superposition of optical variability from different regions in jets for BL Lacs without
the color-brightness correlations. The BL Lacs with the BWB trends may have a single emitting region of optical variability. The relativistic electrons in a single emitting region can produce the broadband SED containing the synchrotron and IC components \citep[e.g.,][]{Li14}. This single emitting region of optical variability will avoid superposing of optical variability from different regions and weakening of the color-brightness correlations. Thus, the variability of brightness, color, and spectral index is likely caused by the change of the underlying relativistic electron distribution that generates the relevant radiation behavior observed in S5 0716+714 as a shock passes through a high density region in jet. This passing of shock will produce SED's variability, such as SED's shape, peak frequency, and peak intensity.

In order to research short timescale optical variability of $\gamma$-ray blazar S5 0716+714, quasi-simultaneous spectroscopic and multi-band photometric observations were performed from 2018 November to 2019 March with the 2.4 m optical telescope located at Lijiang Observatory of Yunnan Observatories. As the BWB trends are detected in the photometric observations, what will the optical spectra show and how will vary? First, the observed spectra can be well fitted with a power-law $F_{\lambda}=A\lambda ^{-\alpha}$. Then we study $\dot{\alpha}$, $\dot{A}$, $\dot{A}/A$, $\dot{F_{\rm{\lambda}}}$, and $\dot{F_{\rm{\lambda}}}/F_{\rm{\lambda}}$ for spectroscopic observations. We find correlations between these quantities, \textbf{which} are consistent with the BWB trends. Interestingly, $\alpha$ is correlated to $F_{\rm{\lambda}}$ and the variations of $\alpha$ lead those of $F_{\rm{\lambda}}$. \textbf{The variations of $\alpha$ indicate variations of relativistic electron distribution producing these optical spectra.} A correlation among $\dot{\alpha}$, $\dot{A}/A$, and $\dot{F_{\rm{\lambda}}}/F_{\rm{\lambda}}$ is found as well. Colors, magnitudes, color variation rates, and magnitude variation rates are studied for photometric observations. We also find correlations between these quantities, which are consistent with the BWB trends. Moreover, the color variations lead the magnitude variations. The data of spectroscopic and photometric observations are mostly distributed in the I(+,+) and III(-,-) quadrants of coordinate system (see Figures 7--9). The observed BWB may be explained by the shock-jet model. Whether there are BWB trends may depend on the relative locations of the synchrotron peak frequencies with respect to the observational frequency ranges, e.g., at the left of the synchrotron peak frequencies. \textbf{Both of spectroscopic and photometric observations give $F_{\rm{var}}\sim 40\%$ which show violent variations in S5 0716+714. Moreover, the range of $\alpha$ is similar to those of colors computed from magnitudes and this similarity implies the reliability of BWB observed in our observations. There are similarities and differences for BWB observed in the spectroscopic and photometric observations. These differences indicate the bandwidth effect on BWB.}

\acknowledgements{We are grateful to the anonymous referee for constructive comments leading to significant improvement
of this work. Thanks for the helpful comments from the ApJ statistics editor. We thank the financial support of the Key Research Program of the CAS (grant No. KJZD-EW-M06), the National Natural Science Foundation of China (NSFC; grant No. 11433004), and the Ministry of Science and Technology of China (2016YFA0400700). We also thank the financial support of
the NSFC (grant No. 11273052), and the CAS Interdisciplinary Innovation Team. We acknowledge the support of the staff of
the Lijiang 2.4m telescope. Funding for the telescope has been provided by Chinese Academy of Sciences and the People's Government of Yunnan Province.}

\clearpage

\begin{deluxetable}{lcccccclccccc}
  \tablecolumns{13}
  \setlength{\tabcolsep}{5pt}
  \tablewidth{0pc}
  \tablecaption{Observation logs of S5 0716+714}
  \tabletypesize{\scriptsize}
  \tablehead{
  \colhead{Date}                                  &
  \colhead{Spectral}                 &
  \multicolumn{4}{c}{Photometric Exposure (s)}    &
  \colhead{}                                      &
  \colhead{Date}                                  &
  \colhead{Spectral}                 &
  \multicolumn{4}{c}{Photometric Exposure (s)}    \\ \cline{3-6} \cline{10-13}
  \colhead{}                          &
  \colhead{Exposure (s)}                                      &
  \colhead{$B$}                                   &
  \colhead{$V$}                                   &
  \colhead{$R$}                                   &
  \colhead{$I$}                                   &
  \colhead{}                                      &
  \colhead{}                          &
  \colhead{Exposure (s)}                                      &
  \colhead{$B$}                                   &
  \colhead{$V$}                                   &
  \colhead{$R$}                                   &
  \colhead{$I$}

} \startdata
2018-11-29  &     & 30 & 20 & 15 & 15 & & 2019-01-21  & 200 &    &    &    &    \\
2018-12-07  & 120 &    &    &    &    & & 2019-01-24  & 120 & 30 & 20 & 15 & 10 \\
2018-12-10  & 200 & 30 & 20 & 15 & 15 & & 2019-01-25  & 120 & 30 & 20 & 15 & 10 \\
2018-12-12  & 120 & 30 & 20 & 15 & 10 & & 2019-01-27  & 120 & 30 & 20 & 15 & 10 \\
2018-12-13  & 120 & 30 & 20 & 15 & 10 & & 2019-01-31  & 120 & 30 & 20 & 15 & 10 \\
2018-12-15  & 120 & 30 & 20 & 15 & 10 & & 2019-02-02  & 120 & 30 & 20 & 15 & 10 \\
2018-12-16  &     & 30 & 20 & 15 &    & & 2019-02-05  & 120 & 30 & 20 & 15 & 10 \\
2018-12-19  & 120 & 30 & 20 & 15 & 10 & & 2019-02-08  & 120 & 30 & 20 & 15 & 10 \\
2018-12-20  & 120 & 30 & 20 & 15 & 10 & & 2019-02-10  & 120 & 30 & 20 & 15 & 10 \\
2018-12-21  & 120 & 30 & 20 & 15 & 10 & & 2019-02-12  & 120 & 30 & 20 & 15 & 10 \\
2018-12-22  &     & 30 & 20 & 15 & 10 & & 2019-02-15  & 120 & 30 & 20 & 15 & 10 \\
2018-12-23  &     & 30 & 20 & 15 & 10 & & 2019-02-19  & 200 &    &    &    &    \\
2018-12-24  & 120 & 30 & 20 & 15 & 10 & & 2019-02-20  & 200 & 30 & 20 & 15 & 10 \\
2018-12-25  & 120 & 30 & 20 & 15 & 10 & & 2019-02-22  & 200 & 30 & 20 & 15 & 10 \\
2018-12-26  & 120 & 30 & 20 & 15 & 10 & & 2019-02-24  & 300 & 30 & 20 & 15 & 10 \\
2018-12-30  & 120 & 30 & 20 & 15 & 10 & & 2019-02-26  & 300 & 30 & 20 & 15 & 10 \\
2018-12-31  & 120 & 30 & 20 & 15 & 10 & & 2019-03-02  & 300 & 30 & 20 & 15 & 10 \\
2019-01-02  & 120 & 30 & 20 & 15 & 10 & & 2019-03-03  & 180 & 30 & 20 & 15 & 10 \\
2019-01-06  & 120 & 30 & 20 & 15 & 10 & & 2019-03-06  & 180 & 30 &    & 15 & 30 \\
2019-01-10  & 120 & 30 & 20 & 15 & 10 & & 2019-03-09  & 120 & 30 & 20 & 15 & 10 \\
2019-01-11  & 120 & 30 & 20 & 15 & 10 & & 2019-03-11  & 180 & 30 & 20 & 15 & 10 \\
2019-01-13  & 120 & 30 & 20 & 15 & 10 & & 2019-03-13  & 120 & 30 & 20 & 15 & 10 \\
2019-01-16  & 120 & 30 & 20 & 15 & 10 & & 2019-03-14  & 900 & 30 & 20 & 15 & 10 \\
2019-01-18  & 200 &    &    &    &    & & 2019-03-15  & 180 & 30 & 20 & 15 & 10 \\
2019-01-19  & 200 &    &    &    &    & &             &     &    &    &    &    \\
\enddata
\label{Table1}
\end{deluxetable}

\begin{deluxetable}{ccccc}
  \tablecolumns{4}
  \setlength{\tabcolsep}{5pt}
  \tablewidth{0pc}
  \tablecaption{Variability Amplitudes}
  \tabletypesize{\scriptsize}
  \tablehead{
  \colhead{Band}                        &
  \colhead{$F_{\rm{var}}$ (\%)}                     &
  \colhead{}                             &
  \colhead{Bin}                         &
  \colhead{$F_{\rm{var}}$ (\%)}
} \startdata
B & 41.8$\pm$4.6 &  & Bin1 &  41.2$\pm$4.3\\
V & 41.4$\pm$4.6 &  & Bin2 &  40.4$\pm$4.2\\
R & 40.1$\pm$4.4 &  & Bin3 &  39.8$\pm$4.1\\
I & 39.5$\pm$4.4 &  & Bin4 &  39.3$\pm$4.1\\
  &                &  & Bin5 & 39.0$\pm$4.1\\
  &                &  & Bin6 & 37.9$\pm$3.8\\
\enddata
\label{Table2}
\end{deluxetable}

\begin{deluxetable}{cccc}
  \tablecolumns{4}
  \setlength{\tabcolsep}{5pt}
  \tablewidth{0pc}
  \tablecaption{Photometric data in the $B$ band}
  \tabletypesize{\scriptsize}
  \tablehead{
  \colhead{JD}                        &
  \multicolumn{3}{c}{$B$}               \\ \cline{2-4}
  \colhead{}                          &
  \colhead{Object}                    &
  \colhead{}                    &
  \colhead{Star3}
} \startdata
2458452.42872 & 14.563 $\pm$ 0.009 &  & 13.286   \\
2458463.25048 & 14.339 $\pm$ 0.002 &  & 13.295   \\
2458465.23931 & 14.189 $\pm$ 0.002 &  & 13.295   \\
2458466.24818 & 14.047 $\pm$ 0.001 &  & 13.294   \\
2458468.34552 & 14.219 $\pm$ 0.008 &  & 13.302   \\
.... & ... &  & ... \\
\enddata
\tablecomments{\footnotesize  JD: Julian dates; Object: magnitudes and errors of
S5 0716+714; Star3: magnitudes of the comparison star. (This table is available
in its entirety in machine-readable form.)}
\label{Table3}
\end{deluxetable}

\begin{deluxetable}{cccc}
  \tablecolumns{4}
  \setlength{\tabcolsep}{5pt}
  \tablewidth{0pc}
  \tablecaption{Photometric data in the $V$ band}
  \tabletypesize{\scriptsize}
  \tablehead{
  \colhead{JD}                        &
  \multicolumn{3}{c}{$V$}               \\ \cline{2-4}
  \colhead{}                          &
  \colhead{Object}                    &
  \colhead{}                    &
  \colhead{Star3}
} \startdata
2458452.42915 & 13.796 $\pm$ 0.014 &  & 12.424   \\
2458463.25093 & 13.529 $\pm$ 0.002 &  & 12.438   \\
2458465.23978 & 13.393 $\pm$ 0.001 &  & 12.437   \\
2458466.24862 & 13.258 $\pm$ 0.001 &  & 12.437   \\
2458468.34611 & 13.402 $\pm$ 0.005 &  & 12.442   \\
... & ... &  & ... \\
\enddata
\tablecomments{\footnotesize  The meaning of each column is same with Table 3. (This
table is available in its entirety in machine-readable form.)}
\label{Table4}
\end{deluxetable}

\begin{deluxetable}{cccc}
  \tablecolumns{4}
  \setlength{\tabcolsep}{5pt}
  \tablewidth{0pc}
  \tablecaption{Photometric data in the $R$ band}
  \tabletypesize{\scriptsize}
  \tablehead{
  \colhead{JD}                        &
  \multicolumn{3}{c}{$R$}               \\ \cline{2-4}
  \colhead{}                          &
  \colhead{Object}                    &
  \colhead{}                    &
  \colhead{Star3}
} \startdata
2458452.42950 & 13.396 $\pm$ 0.011 &  & 12.056   \\
2458463.25130 & 13.089 $\pm$ 0.002 &  & 12.064   \\
2458465.24019 & 12.967 $\pm$ 0.003 &  & 12.064   \\
2458466.24899 & 12.849 $\pm$ 0.005 &  & 12.062   \\
2458468.34655 & 12.965 $\pm$ 0.004 &  & 12.063   \\
... & ... &  & ... \\
\enddata
\tablecomments{\footnotesize  The meaning of each column is same with Table 3. (This
table is available in its entirety in machine-readable form.)}
\label{Table5}
\end{deluxetable}

\begin{deluxetable}{cccc}
  \tablecolumns{4}
  \setlength{\tabcolsep}{5pt}
  \tablewidth{0pc}
  \tablecaption{Photometric data in the $I$ band}
  \tabletypesize{\scriptsize}
  \tablehead{
  \colhead{JD}                        &
  \multicolumn{3}{c}{$I$}               \\ \cline{2-4}
  \colhead{}                          &
  \colhead{Object}                    &
  \colhead{}                    &
  \colhead{Star3}
} \startdata
2458452.42980 & 12.911 $\pm$ 0.008 &  & 11.779   \\
2458463.25164 & 12.536 $\pm$ 0.025 &  & 11.762   \\
2458465.24049 & 12.459 $\pm$ 0.001 &  & 11.787   \\
2458466.24932 & 12.352 $\pm$ 0.002 &  & 11.788   \\
2458468.34704 & 12.437 $\pm$ 0.004 &  & 11.786   \\
... & ... &  & ... \\
\enddata
\tablecomments{\footnotesize  The meaning of each column is same with Table 3. (This
table is available in its entirety in machine-readable form.)}
\label{Table6}
\end{deluxetable}

\begin{deluxetable}{ccccccc}
  \tablecolumns{7}
  \setlength{\tabcolsep}{5pt}
  \tablewidth{0pc}
  \tablecaption{Spectral flux in each bin.}
  \tabletypesize{\scriptsize}
  \tablehead{
  \colhead{JD}                        &
  \colhead{Bin1}                      &
  \colhead{Bin2}                      &
  \colhead{Bin3}                      &
  \colhead{Bin4}                      &
  \colhead{Bin5}                      &
  \colhead{Bin6}
} \startdata
2458460.226840 & 1.361 $\pm$ 0.036 & 1.224 $\pm$ 0.036 & 1.136 $\pm$ 0.020 & 1.030 $\pm$ 0.027 & 0.988 $\pm$ 0.023 & 0.892 $\pm$ 0.024 \\
2458463.255093 & 1.682 $\pm$ 0.019 & 1.525 $\pm$ 0.022 & 1.395 $\pm$ 0.024 & 1.280 $\pm$ 0.018 & 1.228 $\pm$ 0.029 & 1.117 $\pm$ 0.023 \\
2458465.244306 & 1.928 $\pm$ 0.041 & 1.677 $\pm$ 0.023 & 1.550 $\pm$ 0.024 & 1.408 $\pm$ 0.027 & 1.330 $\pm$ 0.029 & 1.200 $\pm$ 0.035 \\
2458466.257454 & 2.277 $\pm$ 0.058 & 1.953 $\pm$ 0.029 & 1.798 $\pm$ 0.030 & 1.610 $\pm$ 0.027 & 1.521 $\pm$ 0.037 & 1.397 $\pm$ 0.037 \\
2458468.356979 & 1.922 $\pm$ 0.043 & 1.705 $\pm$ 0.031 & 1.570 $\pm$ 0.025 & 1.453 $\pm$ 0.033 & 1.347 $\pm$ 0.025 & 1.233 $\pm$ 0.028 \\
... & ... & ... & ... & ... & ... & ... \\
\enddata
\tablecomments{\footnotesize  (This table is available in its entirety in
machine-readable form.)}
\label{Table7}
\end{deluxetable}

\begin{deluxetable}{cccccc}
  \tablecolumns{6}
  \setlength{\tabcolsep}{5pt}
  \tablewidth{0pc}
  \tablecaption{Spearman's rank analysis results.}
  \tabletypesize{\scriptsize}
  \tablehead{\colhead{X}  &  \colhead{Y} &  \colhead{$r_{\rm{s}}$} &  \colhead{$P_{\rm{s}}$}
  &  \colhead{$r_{\rm{s}}$(MC)} &  \colhead{$-\log P_{\rm{s}}$(MC)}
} \startdata

 $\dot{\alpha}$ & $\dot{F_{\lambda}}$ & 0.800 &  $< 10^{-4}$ &  0.70$\pm$0.05  &  7.3$\pm$1.5  \\

 $\dot{A}$  &$\dot{F_{\lambda}}$  & 0.864  & $< 10^{-4}$  &   0.79$\pm$0.05   &  10.5$\pm$1.9   \\

 $\dot{\alpha}$ & $\dot{A}$ & 0.856  & $< 10^{-4}$  & 0.76$\pm$0.05   & 9.4$\pm$1.8  \\

$\dot{\alpha}$ &$\dot{F_{\lambda}}/F_{\lambda}$ & 0.787  & $< 10^{-4}$  & 0.69$\pm$0.06 & 7.2$\pm$1.5  \\

$\dot{A}/A$ & $\dot{F_{\lambda}}/F_{\lambda}$ & 0.875 &   $< 10^{-4}$ & 0.77$\pm$0.05 & 9.8$\pm$2.0  \\

 $\dot{\alpha}$ & $\dot{A}/A$ & 0.971  & $< 10^{-4}$  & 0.85$\pm$0.05 & 13.4$\pm$2.8 \\

 $\Delta B/ \Delta T$  &  $ \Delta(B-I)/ \Delta T$ & 0.781  &  $< 10^{-4}$  & 0.78$\pm$0.02 &  8.6$\pm$0.8  \\

 $B$    &  $B - I$ & 0.553  & 2 $\times 10^{-4}$  & 0.54$\pm$0.02 &  3.6$\pm$0.2 \\

 $B$    &  $B - V$ & 0.395  & 1 $\times 10^{-2}$  & 0.41$\pm$0.06 &  2.2$\pm$0.6 \\

 $B$    &  $V - R$ & 0.470  & 2 $\times 10^{-3}$  & 0.44$\pm$0.06 &  2.5$\pm$0.6 \\

 $B$    &  $R - I$ & 0.492  & 1 $\times 10^{-3}$  & 0.49$\pm$0.04 &  3.0$\pm$0.5 \\

\enddata
\tablecomments{\footnotesize X and Y are the relevant quantities of spectra fitted in section 2 and \textbf{these presented in Figures 6 and 8.}}
\label{Table8}
\end{deluxetable}

\begin{figure*}
  \includegraphics[scale = 0.8]{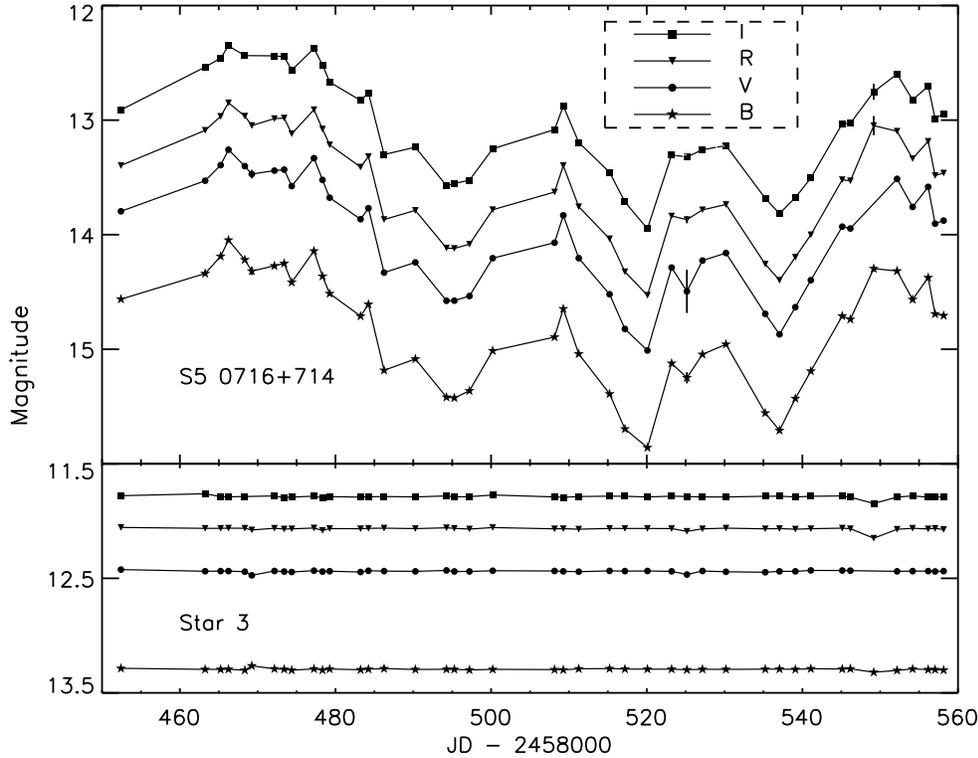}
 \caption{Photometric LCs: the upper panel represents the light curves of S5 0716+714, the lower panel denotes the corresponding variations of comparison star.}
  \label{fig1}
\end{figure*}

\clearpage

\begin{figure*}
 \begin{center}
  \includegraphics[angle=0,scale=0.8]{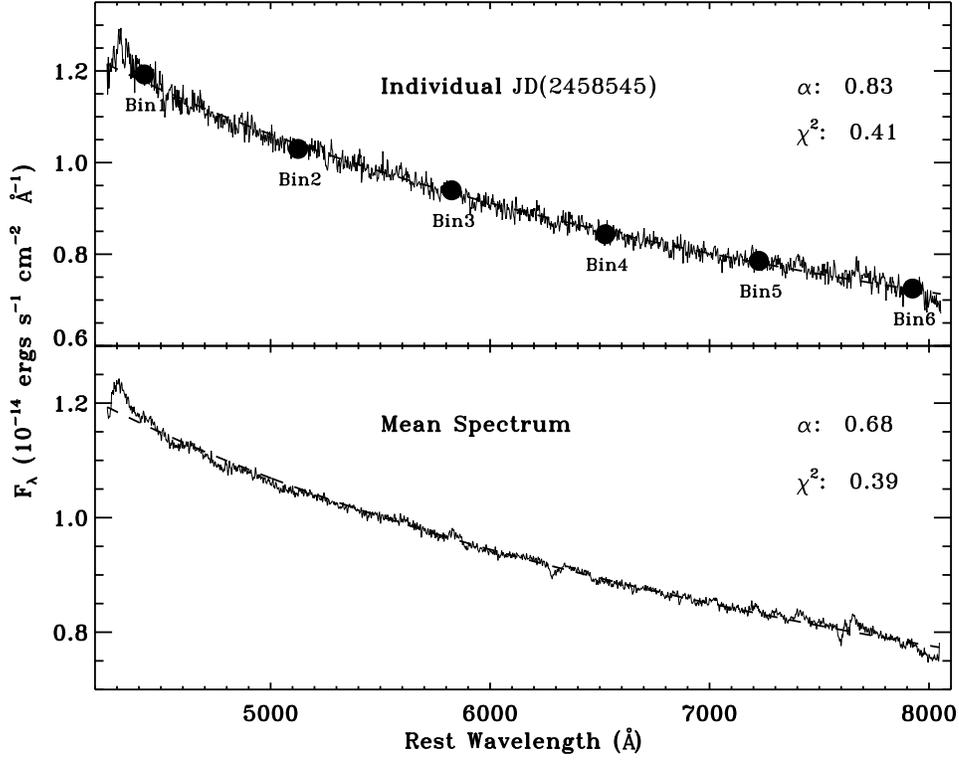}
 \end{center}
 \caption{Individual spectrum (top panel), and the mean spectrum (bottom panel) of S5 0716+714. In each panel, the solid and
  dash lines represent the observed spectrum and the relevant best fitting, respectively. The circles denote the average flux density in each bin of the spectrum.}
  \label{fig2}
\end{figure*}

\clearpage

\begin{figure*}
  \includegraphics[scale=0.8]{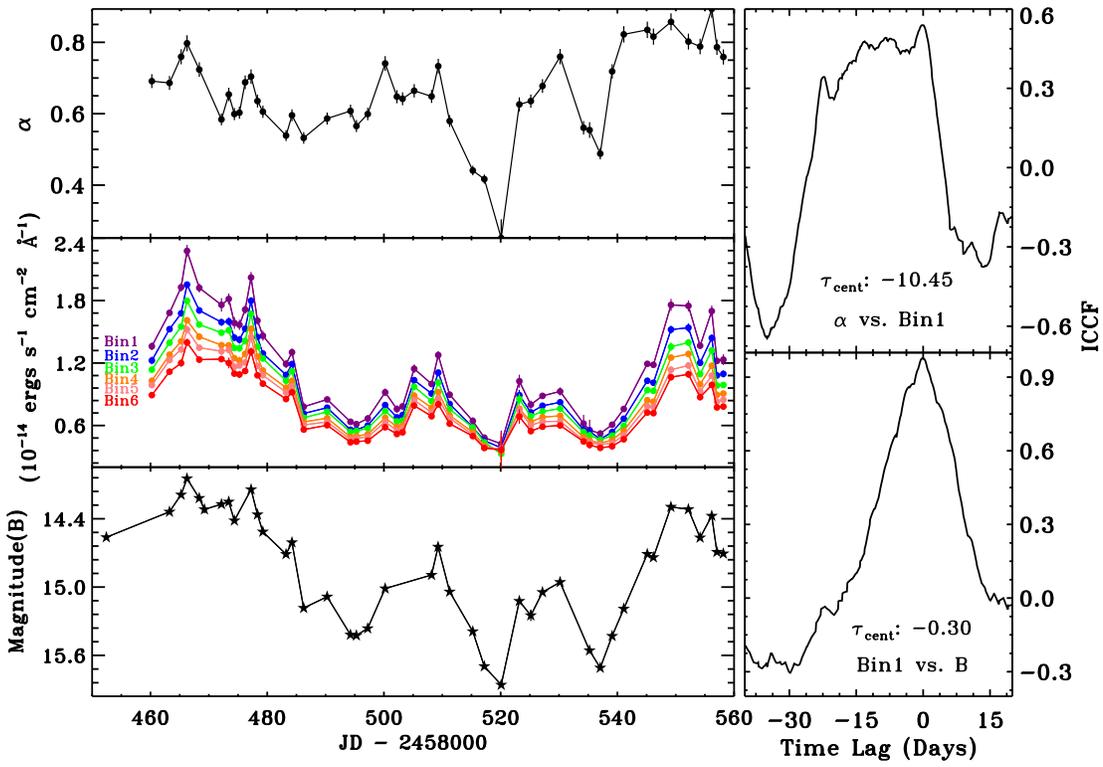}
 \caption{The left panel shows the variability of spectral index (top), flux densities within different bins (middle),
  and $B$ magnitude (bottom). The right panel shows two interpolation cross-correlation functions.}
  \label{fig3}
\end{figure*}

\begin{figure*}
  \includegraphics[scale = 0.8]{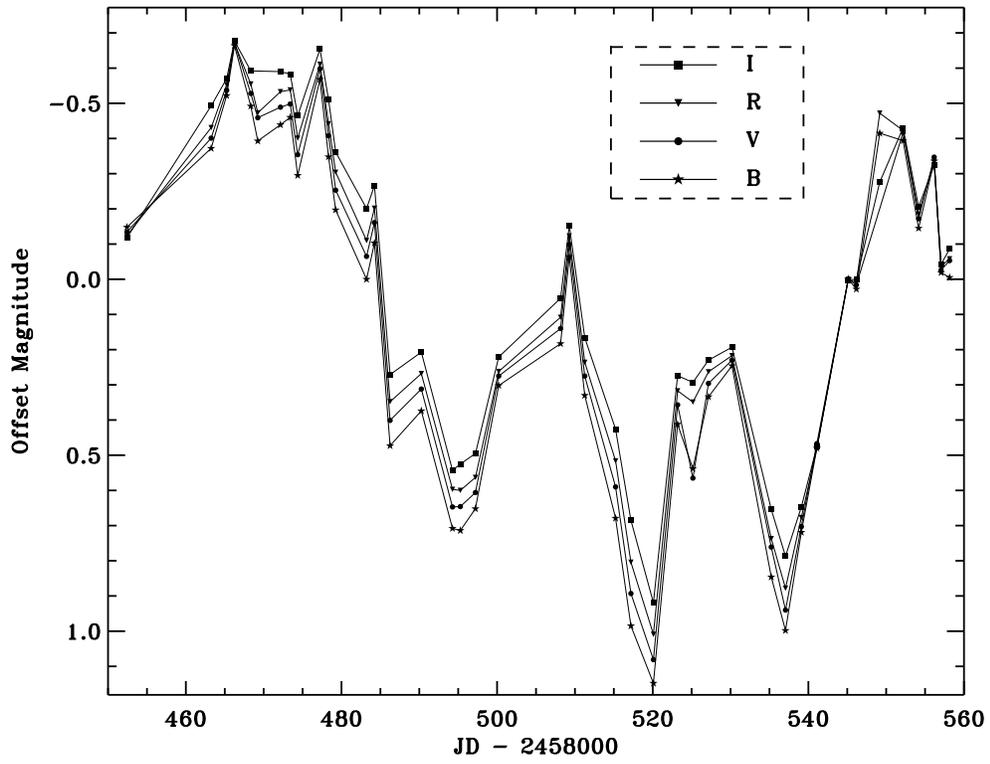}
 \caption{Offset photomectric light curves of S5 0716+714.}
  \label{fig4}
\end{figure*}

\begin{figure*}
  \includegraphics[scale = 0.8]{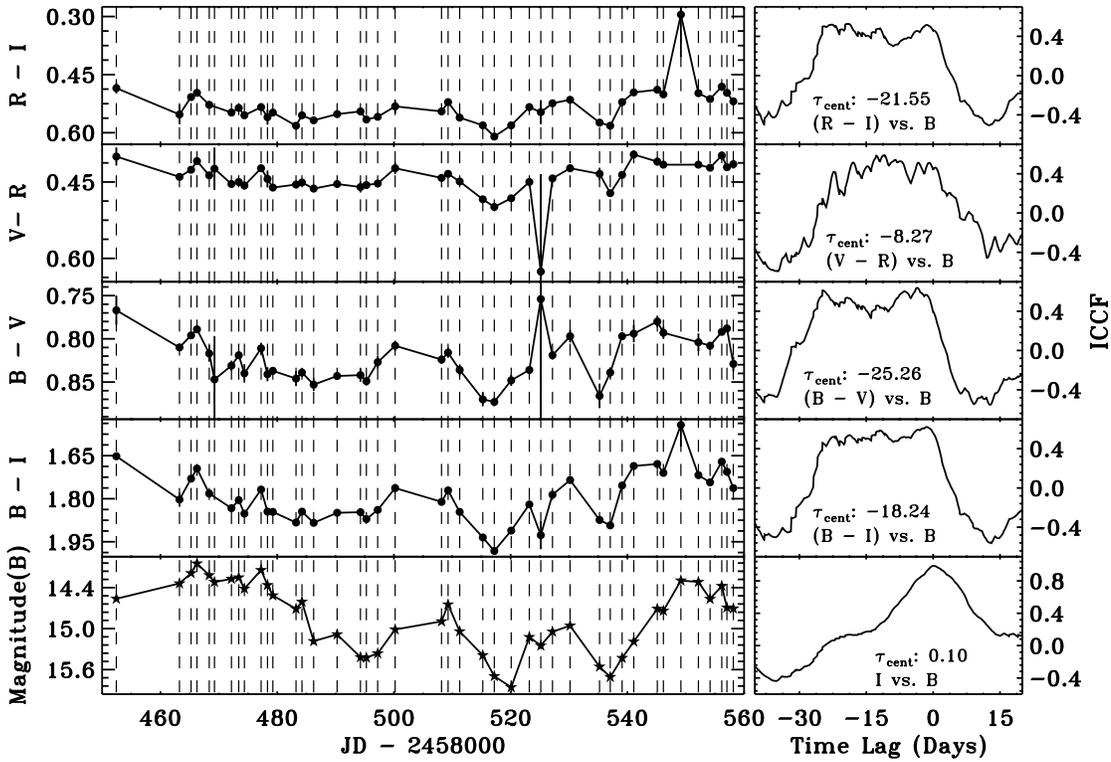}
 \caption{The left panel shows colour–-magnitude diagrams of S5 0716+714. The right panel shows interpolation
  cross-correlation functions of colour vs magnitude, and $I$ and $B$ LCs.}
  \label{fig5}
\end{figure*}

\begin{figure*}
  \includegraphics[scale = 0.8]{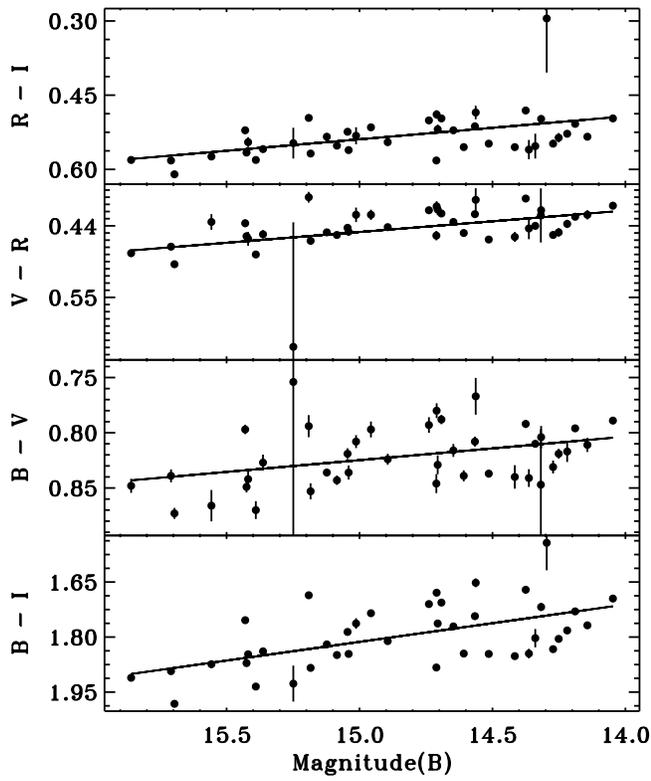}
 \caption{Colour–-magnitude diagrams of S5 0716+714. In each panel, the solid line is the best fitting.}
  \label{fig6}
\end{figure*}

\begin{figure*}
 \begin{center}
  \includegraphics[angle=-90,scale=0.45]{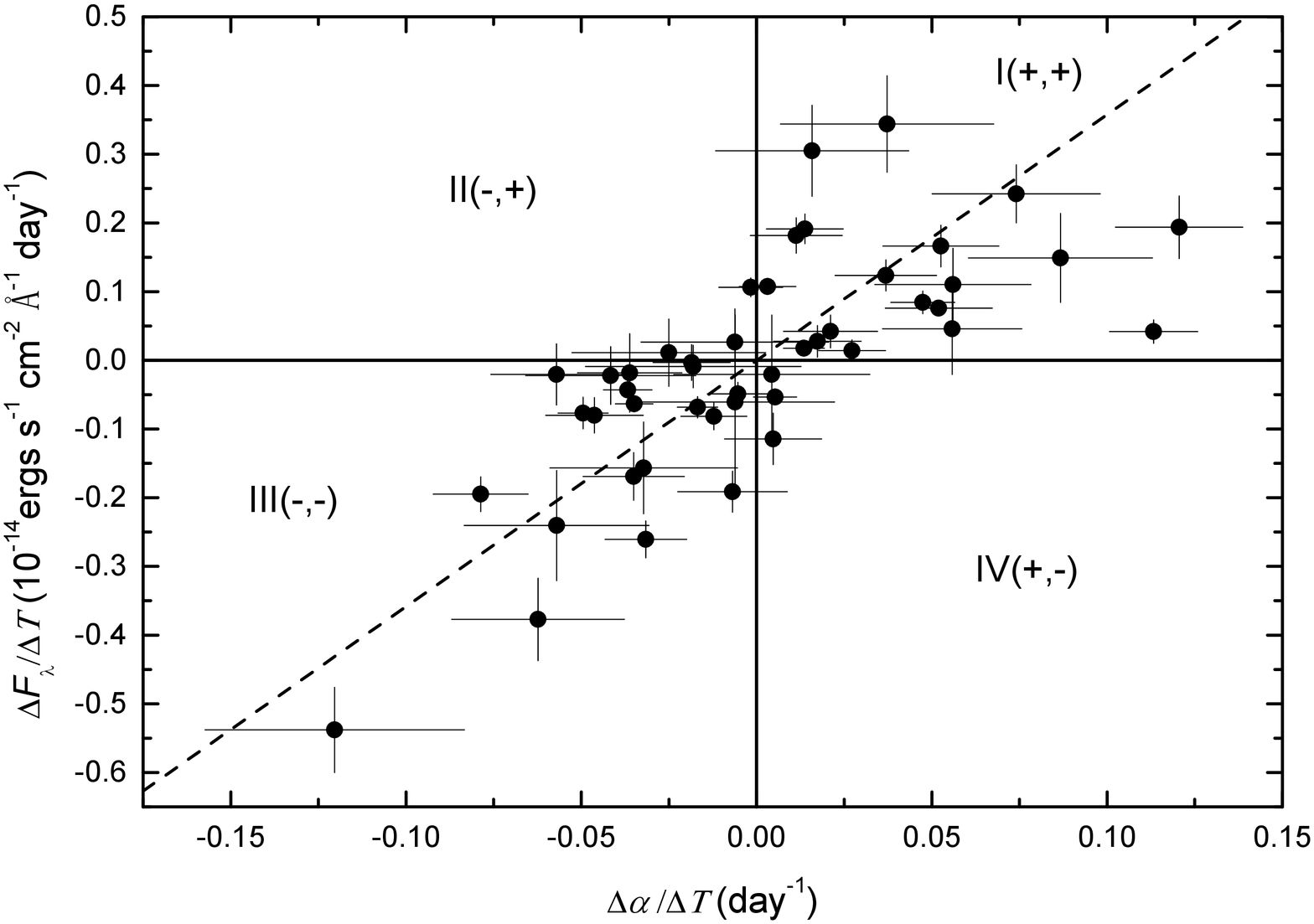}
 \end{center}
 \caption{$\dot{F_{\lambda}}$ vs. $\dot{\alpha}$ within Bin1. The FITEXY estimator \citep{Pr92} gives the best linear fitting [$y=-4(\pm 70)\times 10^{-4}+3.58(\pm0.25)\times x$]. I(+,+), III(-,-), II(-,+), and IV(+,-) are four quadrants of coordinate system.}
\end{figure*}

\begin{figure*}
 \begin{center}
  \includegraphics[angle=-90,scale=0.45]{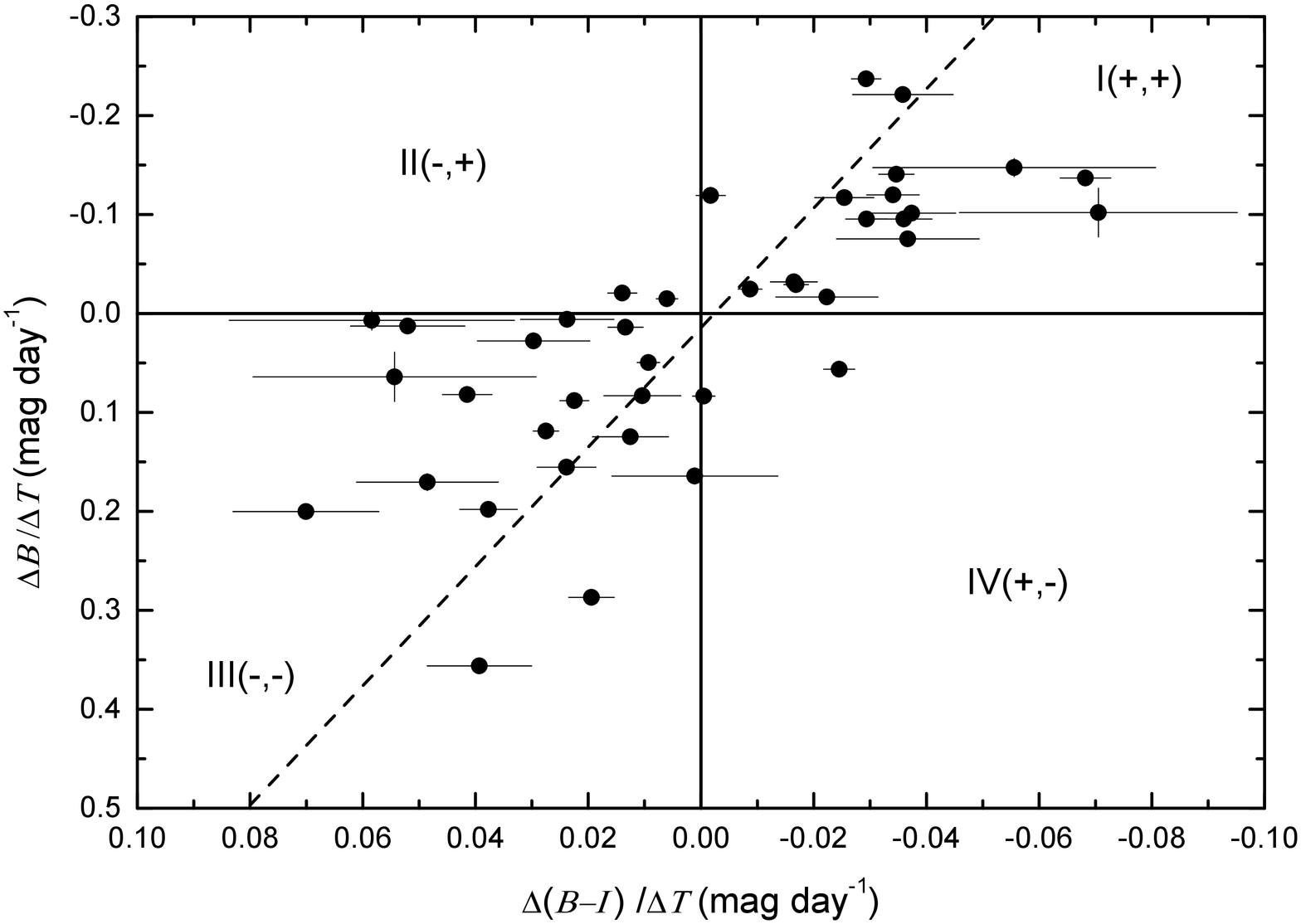}
 \end{center}
 \caption{$\Delta B/\Delta T$ vs. $\Delta(B - I)/\Delta T$. The FITEXY estimator gives the best linear fitting [$y=1.4(\pm 0.4)\times 10^{-2}+6.03(\pm0.20)\times x$].}
  \label{fig8}
\end{figure*}

\begin{figure*}
 \begin{center}
  \includegraphics[angle=-90,scale=0.45]{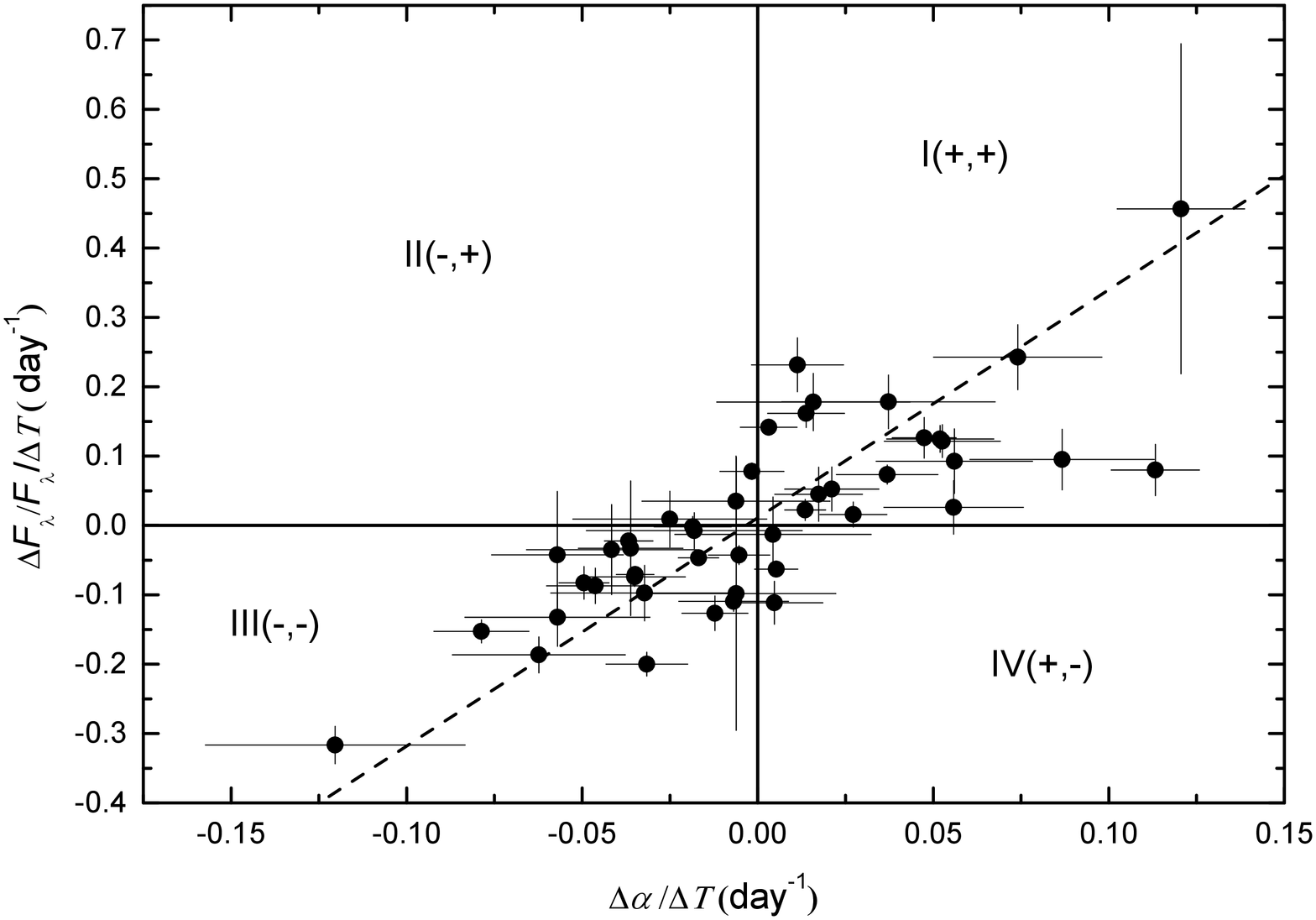}
 \end{center}
 \caption{$\dot{F_{\lambda}}/F_{\lambda}$ vs. $\dot{\alpha}$ within Bin1. The FITEXY estimator gives the best linear fitting [$y=1.1(\pm0.7)\times 10^{-2}+3.29(\pm0.23)\times x$].}
  \label{fig9}
\end{figure*}

\begin{figure*}
 \begin{center}
  \includegraphics[angle=-90,scale=0.45]{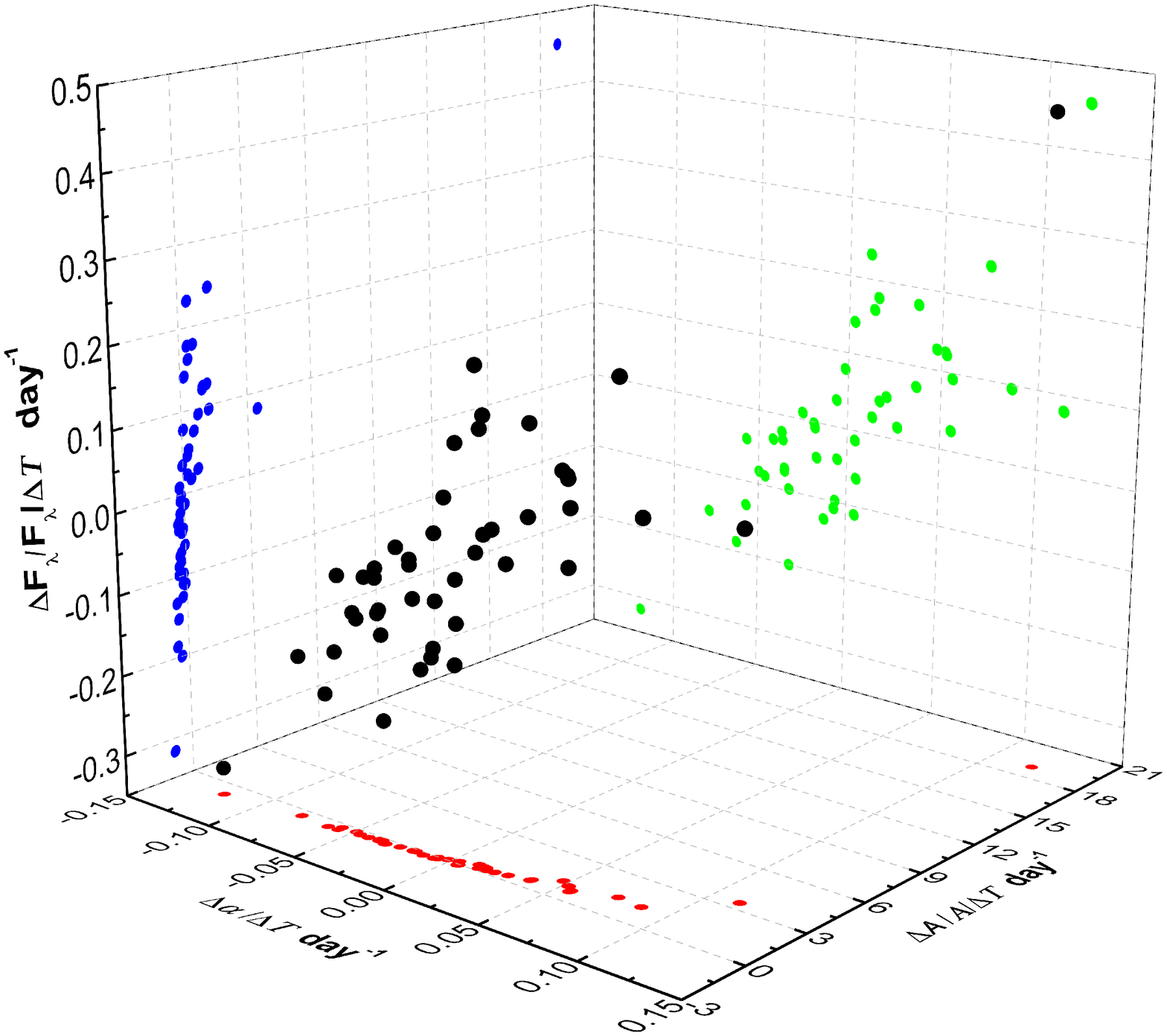}
 \end{center}
 \caption{(X,Y,Z)=$(\dot{\alpha},\dot{A}/A,\dot{F_{\lambda}}/F_{\lambda})$ with $F_{\lambda}$ within Bin1. Color points correspond to XY, XZ, and YZ projections of black points.}
  \label{fig10}
\end{figure*}

\begin{figure*}
  \includegraphics[scale = 0.8]{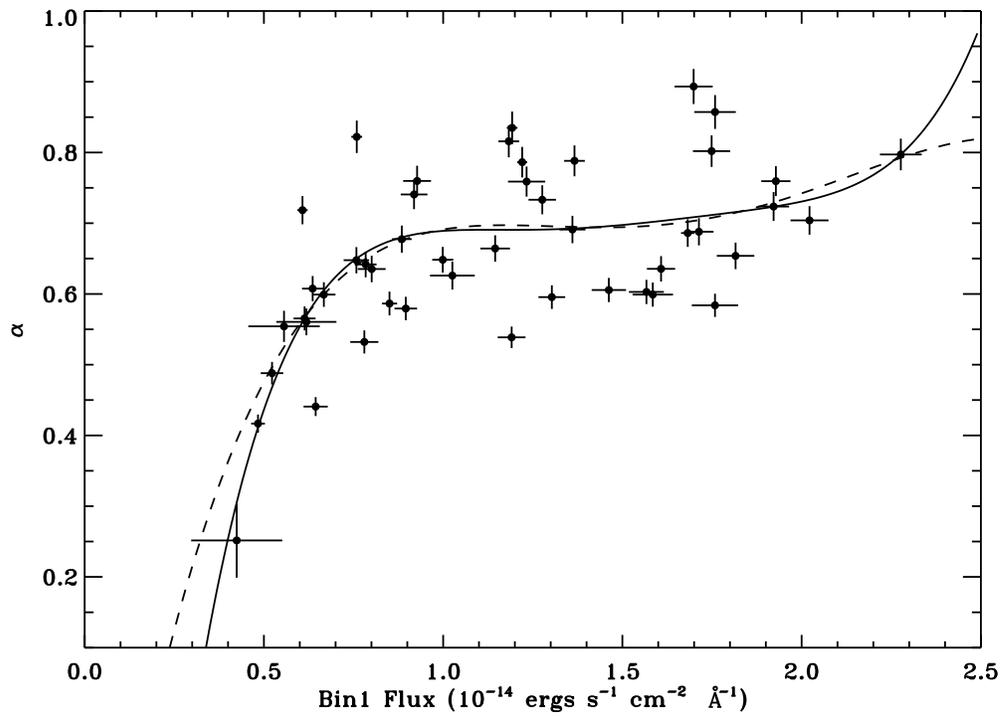}
 \caption{Spectral index vs. flux density within Bin1. The solid and dash lines are the best fitting for all the data and these data excluding the lowest point, respectively.}
  \label{fig10}
\end{figure*}

\end{document}